\documentclass[prb,twocolumn,superscriptaddress]{revtex4-2}
\usepackage{graphicx}
\usepackage{bm}
\usepackage{amsmath,amssymb}
\usepackage{color}
\usepackage{gensymb}

\definecolor{darkGreen}{RGB}{0,110,0}
\definecolor{darkBlue}{RGB}{0,0,130}
\usepackage[colorlinks,citecolor=darkGreen,linkcolor=darkBlue,urlcolor=blue,hyperindex]{hyperref}

\DeclareMathOperator{\tr}{Tr}
\newcommand{\be}{\begin{equation}}
\newcommand{\ee}{\end{equation}}
\renewcommand{\P}{\mathcal{P}}
\newcommand{\bra}[1]{\langle #1|}
\newcommand{\ket}[1]{|#1\rangle}
\newcommand{\braket}[2]{\langle #1|#2\rangle}
\newcommand{\ketbra}[2]{| #1 \rangle \langle #2|}
\newcommand{\brakett}[3]{\langle #1|#2|#3\rangle}

\newcommand{\id}{\openone}

\begin{document}

\title{Locality optimization for parent Hamiltonians of Tensor Networks}

\author{G.\ Giudici}
\affiliation{\mbox{Arnold Sommerfeld Center for Theoretical Physics, University of Munich, Theresienstra\ss{}e 37, 80333 M\"{u}nchen, Germany}}
\affiliation{Munich Center for Quantum Science and Technology (MCQST), Schellingstra\ss{}e~4, 80799 M\"{u}nchen, Germany}

\author{J.I.\ Cirac}
\affiliation{Munich Center for Quantum Science and Technology (MCQST), Schellingstra\ss{}e~4, 80799 M\"{u}nchen, Germany}
\affiliation{Max-Planck-Institute of Quantum Optics, Hans-Kopfermann-Straße 1, 85748 Garching, Germany}
 
\author{N.\ Schuch}
\affiliation{Munich Center for Quantum Science and Technology (MCQST), Schellingstra\ss{}e~4, 80799 M\"{u}nchen, Germany}
\affiliation{Max-Planck-Institute of Quantum Optics, Hans-Kopfermann-Straße 1, 85748 Garching, Germany}
\affiliation{\mbox{University of Vienna, Faculty of Physics,  Boltzmanngasse
5, 1090 Wien, Austria}}
\affiliation{\mbox{University of Vienna, Faculty of Mathematics, Oskar-Morgenstern-Platz 1, 1090 Wien, Austria}}

\begin{abstract}
Tensor Network states form a powerful framework for both the analytical and numerical study of strongly correlated phases. Vital to their analytical utility is that they appear as the exact ground states of associated parent Hamiltonians, where canonical proof techniques guarantee a controlled ground space structure.  Yet, while those Hamiltonians are local by construction, the known techniques often yield complex Hamiltonians which act on a rather large number of spins. 
In this paper, we present an algorithm to systematically simplify parent Hamiltonians, breaking them down into any given basis of elementary interaction terms.  The underlying optimization problem is a semidefinite program, and thus the optimal solution can be found efficiently. Our method exploits a degree of freedom in the construction of parent Hamiltonians -- the excitation spectrum of the local terms -- over which it optimizes such as to obtain the best possible approximation.
We benchmark our method on the AKLT model and the Toric Code model, where we show that the canonical parent Hamiltonians (acting on 3 or 4 and 12 sites, respectively) can be broken down to the known optimal 2-body and 4-body terms. We then apply our method to the paradigmatic
Resonating Valence Bond (RVB) model on the kagome lattice. Here,
the simplest previously known parent Hamiltonian acts on all the 12 spins on one kagome star. With our optimization algorithm, we obtain a vastly simpler Hamiltonian: We find that the RVB model is the exact ground state of a parent Hamiltonian whose terms
are all products of at most four Heisenberg interactions, and whose range can be further constrained, providing a major improvement over the previously known 12-body Hamiltonian. 
\end{abstract} 

\maketitle

\section{Introduction}

Tensor Network States (TNS), in particular one-dimensional Matrix Product States (MPS) and higher-dimensional Projected Entangled Pair States
(PEPS), form a powerful framework for the study of strongly correlated
quantum many-body systems.  They describe complex many-body
wavefunctions by associating local tensors to individual sites which are
then correlated locally, based upon the understanding that the
entanglement 
follows the locality of the interactions.
MPS and PEPS therefore provide a faithful approximation of low-energy
states of local Hamiltonians, which -- together with efficient algorithms
for their variational optimization -- makes these states the basis of a
wide variety of numerical
methods~\cite{schollwoeck:review-annphys,bridgeman:interpretive-dance,molnar:thermal-peps,white:DMRG,verstraete:2D-dmrg,jordan:iPEPS,vanderstraeten:iPEPS-gradient}.

At the same time,  MPS and PEPS also provide an effective toolkit for the
{\it analytical} study of quantum many-body systems~\cite{cirac:tn-review-2021},
for several reasons: First, Tensor Networks allow to model global
symmetries locally, enabling one to understand their action on the
entanglement~\cite{sanz:mps-syms,perez-garcia:inj-peps-syms,molnar:normal-peps-fundamentalthm};
second, many interesting wavefunctions have an exact MPS or PEPS
representation (such as topological fixed point models or Anderson's
Resonating Valence Bond
wavefunction)~\cite{verstraete:comp-power-of-peps,schuch:rvb-kagome,buerschaper:stringnet-peps,gu:stringnet-peps};
and third, any MPS or PEPS appears as the exact ground state of some
associated local \emph{parent
Hamiltonian}~\cite{fannes:FCS,perez-garcia:mps-reps,perez-garcia:parent-ham-2d,molnar:normal-peps-fundamentalthm,schuch:peps-sym,buerschaper:twisted-injectivity,sahinoglu:mpo-injectivity}.
Importantly, these Hamiltonians inherit the symmetries encoded in the
tensor, and techniques for lower bounding their gaps have been
established, which makes them perfectly suited for the analytical
characterization of the physics of strongly correlated phases. The most
well-known among such models is arguably the AKLT
(Affleck-Kennedy-Lieb-Tasaki)
model~\cite{affleck:aklt-prl,affleck:aklt-cmp}, which provided the first example of a
spin-$1$ model with $\mathrm{SO(3)}$ symmetry for which the Haldane gap
could be rigorously proven, and which historically gave rise to the
development of MPS as an analytical tool~\cite{cirac:tn-review-2021}.

In constructing parent Hamiltonians, two main desiderata must be met: They
should have a well-behaved ground space, and they should be simple. By
construction, they enforce local consistency of the ground space with the
tensor network description, and a variety of conditions has been derived
which guarantee that this local consistency implies global consistency,
that is, a global ground space which is either unique or has a controlled
degeneracy (such as for topological
phases)~\cite{perez-garcia:mps-reps,perez-garcia:parent-ham-2d,molnar:normal-peps-fundamentalthm,schuch:peps-sym,buerschaper:twisted-injectivity,sahinoglu:mpo-injectivity}.
At the same time, the construction principle of enforcing local
consistency automatically yields a Hamiltonian which is local, i.e., a sum of
local terms. However, though local, these Hamiltonians can still act on
fairly large clusters, in particular in 2D: This is due to the fact that
generically, non-trivial constraints only arise once the degrees of
freedom in the bulk exceed those at the boundary, which requires larger
regions in higher dimensions.  

Remarkably, however, in many cases of practical interest much smaller
Hamiltonians than those derived through canonical (i.e., generally
applicable) proof techniques suffice. For instance, in the 1D AKLT model,
the canonical $3$-body Hamiltonian can be broken down into $2$-body
Hamiltonians, as well as in the 2D honeycomb AKLT model ($10$-body to
$2$-body) or in the Toric Code model ($12$-body to $4$-body). Clearly,
obtaining such significantly simplified Hamiltonians is highly desirable.
Unfortunately,
no systematic procedure for breaking down canonically constructed parent
Hamiltonians is known: To start with, this requires a suitable guess for a simpler
Hamiltonian, which is not always available, and said guess must
subsequently be shown to have the same ground space as the canonical
Hamiltonian, which must either be done brute-force on a computer, or by
proving it by hand on a case-by-case basis.

In this paper, we present a systematic method to break down any given parent
Hamiltonian into sums of simple elementary Hamiltonian terms. The method
can be combined with any canonical tensor network parent Hamiltonian
construction from the literature, and for any given target set of desired
simple Hamiltonian terms. Our method gives rise to an optimization problem
which can be solved systematically and efficiently; formally, the
optimization can be rephrased as a semidefinite program (SDP) and is thus
converging provably efficiently.  A central ingredient which we exploit in this
optimization is an ambiguity in the canonical construction of parent
Hamiltonians, relating to the spectrum of local excitations.

To test our method, we first benchmark it on two well-studied 
models: First, the 1D AKLT model, where we show that
it allows to reduce the $4$-body parent Hamiltonian obtained through the
most fundamental canonical technique to a sum of nearest-neighbor
$2$-body interactions. Second, the Toric Code model, where
more refined canonical techniques allow to construct a $12$-body
Hamiltonian, which our method breaks down to the well-known sum of
$4$-body vertex and plaquette terms. 

We then apply our algorithm to one of the most paradigmatic models of a topological spin liquid, the spin-$1/2$ Resonating Valence Bond (RVB) state on the kagome
lattice. There, our method allows us to obtain a vastly simpler parent Hamiltonian than the ones previously known. Namely, 
for this model, refined canonical techniques gave rise to a $19$-body Hamiltonian
(acting on two overlapping stars)~\cite{schuch:rvb-kagome}, which had subsequently be shown to be a
sum of two $12$-body terms (each acting on one star) by a suitable guess,
confirmed by brute-force numerical
analysis~\cite{zhou:rvb-parent-onestar}. Applying our method to
systematically study the possibility to further decompose this $12$-body
interaction, we find that it can be decomposed into terms containing product of at most four Heisenberg interactions each, thereby significantly reducing the complexity of the parent Hamiltonian of the kagome RVB model.

\begin{figure}
\includegraphics[scale=0.5]{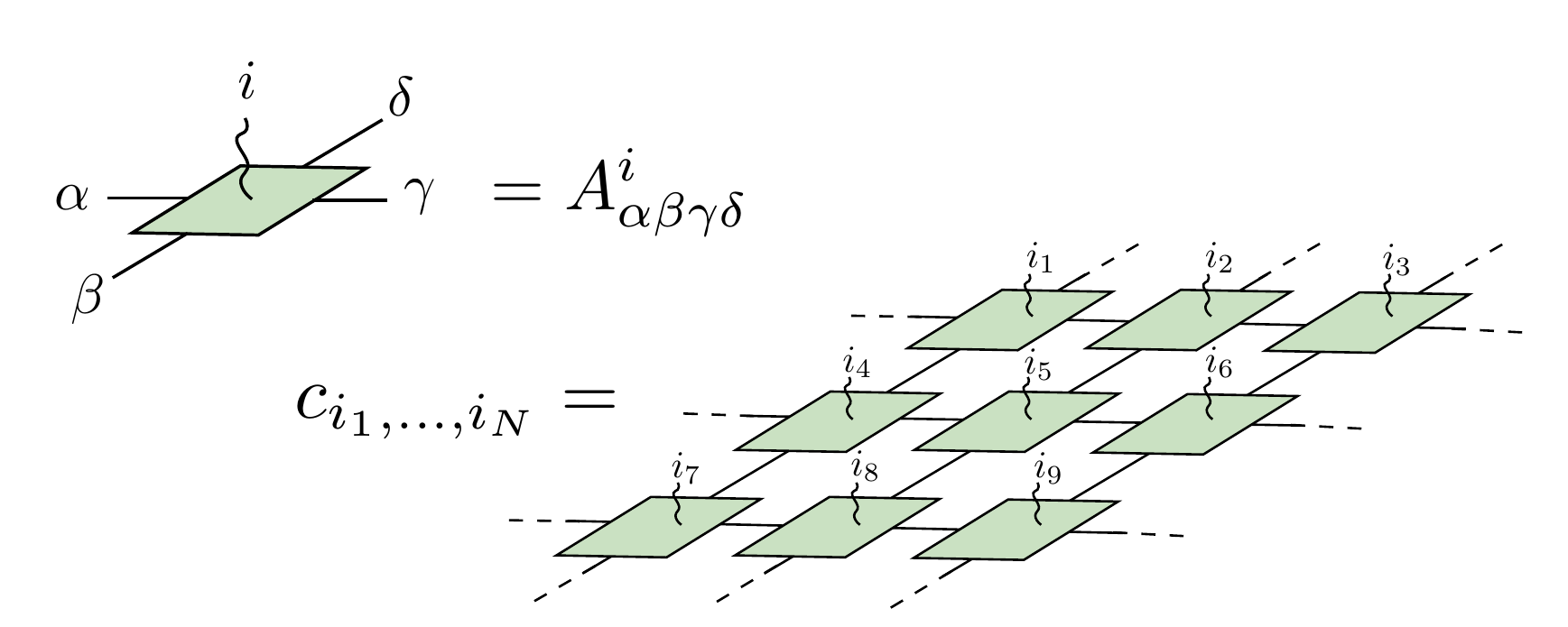}
\caption{Tensor networks (shown for a 2D square lattice) are constucted from a local tensors $A^i_{\alpha\beta\gamma\delta}$ (with physical index $i$ and virtual indices $\alpha,\beta,\ldots$); they are arranged on a lattice and the virtual indices are contracted to yield the expansion coefficient $\ket\Psi = \sum c_{i_1\,\dots,i_N}\ket{i_1,\dots,i_N}$.}
\label{fig:peps}
\end{figure}

\section{Tensor Networks and Parent Hamiltonians}

We start by introducing Tensor Network states and parent Hamiltonians, and in
particular canonical constructions for the latter.

Tensor
Networks~\cite{schollwoeck:review-annphys,bridgeman:interpretive-dance,cirac:tn-review-2021}
are constructed by associating a tensor $A^i_{\alpha\beta\cdots}$ to each
site (we focus on translational
invariant systems, where all tensors are the same), where $i=1,\dots, d$ is the
\emph{physical index} and $\alpha,\beta,\ldots = 1,\dots, D$ are \emph{virtual indices}
or \emph{entanglement indices} with \emph{bond dimension} $D$. The tensors
are then arranged on a lattice and the virtual indices contracted 
(that is, identified and summed over)
with those of the adjacent tensors. Two
particularly important classes are MPS, where the tensors are arranged on
a 1D line, and PEPS, where the tensors are placed on some 2D lattice.
After contracting all indices with suitable boundary conditions (for the
purpose of this work, the boundary conditions are irrelevant, but take
periodic), one is left with a multi-index tensor $c_{i_1,\dots,i_N}$ which
depends on the physical index $i$ of all $N$ spins in the system, and
which provides an MPS or PEPS description of its wavefunction $\ket\Psi$ by virtue of
$\ket\Psi = \sum c_{i_1,\dots,i_N}\ket{i_1,\dots, i_N}$.
The construction can be modified to include more than one type of
tensors, such as tensors which only carry virtual indices and which
are placed in between the tensors with physical indices; we will encounter such
an instance later in Sec.~\ref{sec:tcode}.

 \begin{figure}
 \centering
 \includegraphics[scale=0.67]{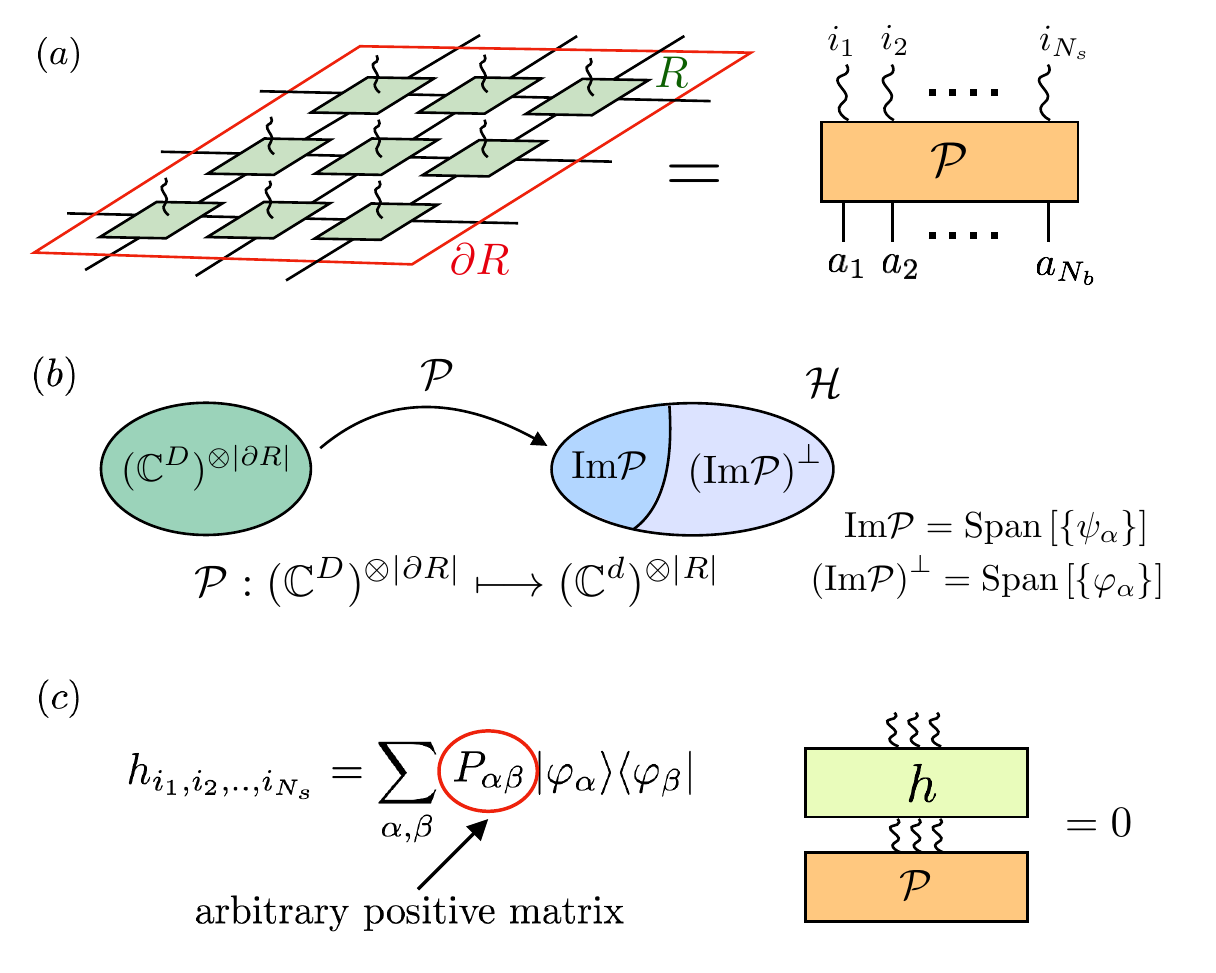} 
 \caption{Schematics of the parent Hamiltonian construction. (a) The elementary tensors are blocked on a region $R$. The resulting tensor is interpreted as a linear map $\mathcal P$ from the virtual to the physical system.  (b) If the blocked region is sufficiently large, the image of the map $\mathcal P$, and thus the support of the overall tensor network wavefunction, will be only a subspace of the full physical Hilbert space $\mathcal H$,  leaving an orthogonal complement $(\mathrm{Im}\,\mathcal P)^\perp$ orthogonal to the state.
(c)~A parent Hamiltonian for the state is obtained by taking an arbitrary positive semi-definite operator $h$ which is zero of $\mathrm{Im}\,\mathcal P$ and strictly positive on $( \mathrm{Im}\, \mathcal{P} )^\perp$. }
 \label{fig:parentham}
 \end{figure}

We now turn to parent Hamiltonians. To this end, consider blocking the
tensors in a  contiguous region $R$, and consider the resulting object as
a linear map $\mathcal P:(\mathbb C^D)^{\otimes |\partial R|}\to(\mathbb
C^d)^{\otimes |R|}$ from
the virtual indices at the boundary $\partial R$ of the region to the
physical indices in the bulk $R$~(Fig.~\ref{fig:parentham}a). Since the volume $|R|$ grows faster than
the boundary $|\partial R|$ as the size of $R$ is (uniformly) increased, one quickly
reaches a point where $\mathrm{Im}\,\mathcal P$ is not full rank~(Fig.~\ref{fig:parentham}b). Then, we
can define a parent Hamiltonian $h\ge0$ such that
$h\big\vert_{\mathrm{Im}\,\mathcal P}=0$ and
$h\big\vert_{\mathcal K}>0$ on $\mathcal K=(\mathrm{Im}\,\mathcal
P)^\perp$ (Fig.~\ref{fig:parentham}c); canonically, one chooses $h$ to be the projector onto $\mathcal K$. This Hamiltonian satisfies $h\ket{\Psi}=0$. If we now take such
an $h_i$ centered around every site $i$ in the lattice, then $H=\sum
h_i\ge 0$,
and $H\ket{\Psi}=\sum h_i\ket\Psi = 0$, that is, the MPS or PEPS
$\ket\Psi$ is a ground state of the \emph{parent Hamiltonian} $H$.

A key question is when this construction gives Hamiltonians with a
well-behaved ground space. For this, the concept of \emph{injectivity} and
the associated \emph{injectivity length} is of central
importance~\cite{cirac:tn-review-2021}.  A
tensor network is said to be \emph{normal} if under blocking sufficiently many
sites, $\mathcal P$
becomes an injective map. The blocked tensor network is then said to be
injective, and the size $L$ of the block (in 1D) is the associated
injectivity length. In 2D, the characterization of the smallest injective
block will depend on the lattice and might not be unique, for the
following, we consider a square region of size $H\times V$ which is
injective.  The relevance
of injectivity lies in the fact that an injective $\mathcal P$ can be
inverted on $\mathrm{Im}\,\mathcal P$ by acting on the physical system.
Thus, an injective tensor network is  equivalent up to local (non-unitary)
transformations to maximally entangled states between nearest neighbors.
The latter is the unique ground state of a two-body Hamiltonian, and by
conjugating this Hamiltonian with the inverse of $\mathcal P$, one arrives
at the conclusion that the two-body parent Hamiltonian constructed from
the blocked tensor network has a unique ground
state~\cite{cirac:tn-review-2021,schuch:mps-phases,darmawan:graph-states-as-ground-states}.
Thus, the parent Hamiltonian acting on $2L$ sites (in 1D) or $2H\times V$
and $H\times 2V$ sites (in 2D) has a unique ground state. This result can
be improved by using more refined proof techniques, which allow to show
that the parent Hamiltonian constructed on $L+1$ (in 1D) or $(H+1)\times
V$ and $H\times (V+1)$ sites (in 2D) has a unique ground
state~\cite{cirac:tn-review-2021,perez-garcia:mps-reps,molnar:normal-peps-fundamentalthm}.

Similar results have been derived for 2D systems exhibiting topological
order: There, the individual PEPS tensors must exhibit an entanglement
symmetry, characterized by a group action or set of Matrix Product
Operators (MPOs), and the relevant scale is set by the block size at which
$\mathcal P$ becomes injective on the symmetric subspace of the
entanglement degrees of freedom ($G$-injectivity or
MPO-injectivity)~\cite{schuch:peps-sym,buerschaper:twisted-injectivity,sahinoglu:mpo-injectivity}.
Again, this implies that the PEPS is equivalent to a topological fixed
point model up to local transformations, which can be used to obtain a
parent Hamiltonian with a topological ground space (e.g., acting on
$2H\times 2V$ sites for Kitaev's double models on the square
lattice)~\cite{schuch:rvb-kagome}; yet again, this can be improved using
more refined techniques to smaller regions, such as $(H+1)\times (V+1)$ on
the square lattice, where the size and shape of the minimal region will
depend on the structure of the tensor network at hand (but not on the
specific model)~\cite{cirac:tn-review-2021}.

In the construction of parent Hamiltonians, there remains a degree of
freedom, beyond the size of the region (the $\mathcal P$) from which the Hamiltonian
is constructed. Namely, we require $h\big\vert_{\mathcal K}>0$ exactly on
$\mathcal K = (\mathrm{Im}\,\mathcal P)^{\perp}$, but any such $h$ will be
a valid choice (and moreover, since any two such operators are relatively
bounded, this choice does not affect gappedness of the total Hamiltonian). This degree of freedom
can be utilized to find an $h$ which can be broken down into a sum of
simpler terms. One example where this happens is the AKLT model, where the
sum of two $2$-body Hamiltonian terms has the same ground space as  
the canonical $3$-body Hamiltonian with a suitably chosen
$h\big\vert_{\mathcal K}>0$. In this specific case, a guess for the
$2$-body Hamiltonian can be obtained from the parent Hamiltonian
construction, applied to $2$ sites. 
However, this need not generally be the case; rather, we
generally expect a parent Hamiltonian to be decomposable into a sum of smaller terms
which by themselves are not parent Hamiltonians (i.e., 
positive semi-definite operators which annihilate the state). Thus, the
question arises whether and how the degree of freedom
$h\big\vert_{\mathcal K}>0$ can be systematically exploited to break down parent
Hamiltonians into a sum of simpler terms. This is precisely the question which we will address
in the remainder of the paper.

\section{Locality optimization algorithm}
\label{ph2_2}

As we have just discussed, the parent Hamiltonian, constructed on any given patch, is highly non-unique, as any operator $h\ge0$ with $h\big\vert_{\mathcal K}>0$ on
$\mathcal K = (\mathrm{Im}\,\mathcal P)^{\perp}$, and zero otherwise, will serve that purpose. In the following, we will devise an algorithm which uses this degree of freedom to break down the parent Hamiltonian into a sum of simpler ``target'' terms.

\subsection{Algorithm\label{sec:algorithm}}

The goal is to expand the target Hamiltonian in some given set of (local) operators $O_a$,
\be \label{Hexpansion}
h = \sum_a c_a O_a,
\qquad
O_a = (O_a)^\dagger.
\ee
In order to keep the dimension of $\{O_a\}$ as small as possible, one can e.g.\ use that the parent Hamiltonian inherits all the symmetries of the TN state, and restrict to a set which shares the same symmetries.
The goal of the algorithm is to find a set of coefficients $\{c_a\}$ such that $h$ is a parent Hamiltonian, that is, $h$ vanishes on 
$\mathrm{Im}\, \P $ and it is strictly positive on its orthogonal complement. To achieve this, we  choose a basis $\{\ket{\varphi_\alpha}\}$ of $\mathrm{Im}\,\mathcal P$ and expand
\be
\label{eq:h-sum-P}
h = \sum_{\alpha, \beta } P_{\alpha\beta} \ketbra{\varphi_\alpha}{\varphi_\beta}\ ,
\ee
where we impose that $P \equiv (P_{\alpha\beta})$ is a strictly positive matrix; this guarantees that $h$ has the required property.
Finding a decomposition $\{c_a\}$ of the parent Hamiltonian now amounts to finding zeros of the cost function
\begin{align}
\nonumber
F\left(c_a,P_{\alpha\beta}\right) & = \Bigg\| \sum_a c_a O_a - \sum_{\alpha, \beta } P_{\alpha\beta} \ketbra{\varphi_\alpha}{\varphi_\beta} \Bigg\|^2_2 =  \\[2mm]
  & = X_A^* \, \mathcal{M}_{A B} \, X_B \stackrel{!}{=} 0\ ,
\label{eigvprob2}
\end{align} 
where the vector $X_A = \left( -c_a , P_{\alpha \beta } \right)$ contains both the information on the expansion coefficients $\{c_a\}$ of $h$ in the operator basis $\{O_a\}$, and on the positive matrix $P_{\alpha\beta}$ representing $h$ as a strictly positive operator on $( \mathrm{Im} \P)^\perp$. The norm $\|\cdot\|_2$ is the Frobenius norm.
If we choose the basis $\{\ket{\varphi_\alpha}\}$ to be orthonormal, 
$\braket{\varphi_\alpha}{\varphi_\beta} = \delta_{\alpha \beta}$, the matrix $\mathcal{M}$ becomes sparse and it reads
\begin{align}
\nonumber
& \hspace{10mm} \mathcal{M}_{A B} = \left( \begin{array}{c|c}
            M_{ab} &  R_{a,\alpha \beta} \\[2mm]
            \hline \\[-2mm]
            R_{\gamma \delta , b}  \, & \, \delta_{\alpha \gamma } \delta_{\beta \delta }
        \end{array}
        \right) ,\\[2mm]
& M_{ab} = \tr ( O_a O_b ) ,
\qquad   R_{a,\alpha \beta } = \brakett{ \varphi_\alpha }{O_a}{\varphi_\beta} .
\label{hessian}
\end{align}

By solving the eigenvalue problem \eqref{eigvprob2}, we thus obtain solutions $h=\sum c_a O_a$ which satisfy
\eqref{eq:h-sum-P} and thus vanish on $\mathrm{Im}\,\mathcal P$. However, $h$ is not positive (or even non-negative) on $(\mathrm{Im}\,\mathcal P)^\perp$ unless we impose $P>0$. To obtain solutions which additionally satisfy this positivity condition, we thus choose to 
minimize the quadratic form in Eq.~\eqref{eigvprob2} on the convex space of positive definite matrices $P_{\alpha\beta}$. Hence, the optimized Hamiltonian is given by the coefficients $c_a$ such that
\be
\label{eq:minimization-problem-P-ge-1}
(c_a,P_{\alpha \beta}) = \mathrm{ArgMin} \, F(c_a,P_{\alpha\beta}),
\qquad  P \ge \id ,
\ee
where we imposed that the eigenvalues of $P$ are $\ge 1$ without loss of generality (as this only amounts to a rescaling).

A possible route to solve this optimization problem is to apply a gradient descent algorithm to the cost function $F$, and project onto the desired space at each step. We start from an initial point such that the Hamiltonian $h$ vanishes, i.e. $c_a = 0$ for all $a$, and the initial $P_{\alpha \beta}$ is the identity. Since the cost function is a quadratic form, the gradient can be efficiently computed by simple matrix multiplication. The algorithm is thus as follows:
\begin{enumerate}
    \item $X_\mathrm{in} = \left( 0 , \id \right)$
    \item $X' = X - \eta \, \nabla F ( X )  = X - 2 \eta \, \mathcal{M} \cdot X  $
    \item $X'' = \mathrm{\Pi} (X') $
    \item Repeat from 2.\ until convergence.
\end{enumerate}
The projection $\Pi$ in 3.\ has to enforce the condition $P\ge \id$. This can be achieved by taking the $P_{\alpha \beta}$ components of the vector $X$, and setting to one all the diagonal elements of the triangular part of its Schur decomposition which dropped below one at step 2., i.e.:
\be 
\mathrm{Proj} ( X ) = \begin{pmatrix} c \\[1mm] \mathrm{\Pi}( P ) \end{pmatrix} = \begin{pmatrix} c \\[1mm] Z^\dagger \, \mathrm{\Pi}( T ) \, Z  \end{pmatrix} =  \begin{pmatrix} c \\[1mm] Z^\dagger \, \widetilde{T} \, Z  \end{pmatrix}, 
\ee
where the triangular matrix $\widetilde{T}$ is the same as $T$, but all the diagonal entries that are smaller than one in $T$ are set to one in $\widetilde{T}$.

To monitor the status of the convergence during the minimization we directly compute the cost function at each step. To speed up the convergence we employ an adaptive step size $\eta_n = \langle \delta X_n , \delta X_n  \rangle /  \langle \delta X_n , \delta G_n \rangle$, where $\delta X_n = X_n - X_{n-1}$ and $\delta G_n = G_n - G_{n-1}$ are the point and gradient displacements at the $n$-th step of the optimization~\cite{adaptive_gd}.

\subsection{Symmetries\label{sec:method-symmetries}}

The dimension of the parameter space of the optimization algorithm is $D_O + ( \mathrm{dim}\,(\mathrm{Im}\,\mathcal P)^\perp  )^2 $, where $D_O$ is the dimension of the operator basis $\{O_a\}$. The number of parameters can be reduced if the target Hamiltonian is invariant under some symmetry. In this case, the matrix $P_{\alpha \beta}$ can be decomposed into blocks labeled by the eigenvalues of the symmetry generator:
\be 
P = \bigoplus_\lambda P^\lambda \ .
\ee
Upon proper choice of the basis of local operators, the same is true for all the $O_a$s. The dimension of the parameter space is thus reduced to $D_O + \sum_\lambda ( D_\lambda )^2 $, where $D_\lambda$ is the dimension of the intersection of $(\mathrm{Im}\, \P )^\perp$ with each eigenspace (irrep) of the symmetry generators. The cost function in Eq.~\eqref{eigvprob2} becomes
\begin{align}
\nonumber
F(c_a,P^\lambda_{\alpha \beta })  & = \sum_\lambda \Bigg\| \sum_a c_a O^\lambda_a - \sum_{\alpha, \beta } P^\lambda_{\alpha\beta} \ketbra{\varphi^\lambda_\alpha}{\varphi^\lambda_\beta} \Bigg\|^2 =  \\[2mm]
\nonumber
& =  c_a \tr ( O_a O_b ) \, c_b
 - 2  c_a \sum_\lambda \brakett{\varphi^\lambda_\alpha}{O^\lambda_a}{\varphi^\lambda_\beta} P^\lambda_{\alpha \beta} + \\[2mm]
 & \quad  + \sum_\lambda ( P^\lambda_{\alpha \beta} )^2  =  X_A^* \, \mathcal{M}_{A B} \, X_B,
\label{cost_sym}
\end{align}
where $O^\lambda$ is the restriction of $O$ to the symmetry sector labeled by $\lambda$, and we used the fact that $O_a$ is block diagonal for all $a$. The vector $X_A = \left( -c_a , P^{\lambda_1}_{\alpha \beta } ,  P^{\lambda_2}_{\alpha \beta } , \dots  \right)$, and the matrix $\mathcal{M}$ reads
\begin{equation}
\begin{aligned}
& \hspace{8mm} \mathcal{M}_{A B} = \left( \begin{array}{c c c c c}
            M &  R_{\lambda_1} & 0 & 0 \\[2mm]
           R^\dagger_{\lambda_1} &  \id_{\lambda_1} & R_{\lambda_2} & 0  \\[2mm]
            0 &  R^\dagger_{\lambda_2} & \id_{\lambda_2} & \ddots \\[2mm]
           0 &  0 & \ddots &  \ddots 
        \end{array}
        \right) ,\\[2mm]
        &  M_{ab} = \tr ( O_a O_b ) \ ,
\qquad R_{a,\alpha \beta } = \brakett{ \varphi_\alpha }{O_a}{\varphi_\beta}\ .
\end{aligned}
\label{hessian2}
\end{equation}
An example which we will use in the following is SU(2) symmetry. In this case we take $\lambda = s,S^z$, where $S^z$ is the eigenvalue of the $z$-component of the total angular momentum on the chosen patch of physical sites, $S^z = \sum_i S^z_i$, and $s$ is the quantum number of the total squared angular momentum $S^2 = (S^x)^2 + (S^y)^2 + (S^z)^2$ (with eigenvalues $s (s+1)$). Note that, thanks to $SU(2)$ invariance, the $P_{\alpha\beta}$ blocks are independent of the eigenvalue $S^z$, and the cost function in Eq.~\eqref{cost_sym} becomes
\begin{align}
\nonumber
F\left(c_a,P^\lambda_{\alpha \beta } \right)  & = c_a \tr ( O_a O_b ) \, c_b + \\[2mm]
\nonumber
& \quad  - 2  c_a \sum_s (2 s + 1)  \brakett{\varphi^{s,0}_\alpha}{O^{s,0}_a}{\varphi^{s,0}_\beta} P^s_{\alpha \beta}  + \\[2mm]
& \quad  + \sum_s ( 2 s + 1 ) ( P^s_{\alpha \beta} )^2,
\end{align}
where the factor $(2 s + 1)$ takes into account the multiplicity of the eigenvalue $S^z$ for a given $s$, so that the generators $\ket{ \varphi^{s,S^z}_\alpha }$ need to be computed only in the $S^z=0$ sector. This further reduces the number of variational parameters down to $D_O + \sum_s (D_s)^2$, where $D_s$ is the dimension of the simultaneous eigenspace of $S^2$ and $S^z$ in the $S^z=0$ sector.

\subsection{Formulation as a semidefinite program}

The minimization problem in Eq.~\eqref{eq:minimization-problem-P-ge-1} can be formulated as a semi-definite program (SDP). This implies that the problem can be systematically and efficiently solved using a suitable SDP solver, and makes clear why the gradient method chosen for the optimization in this work performs so well on the problem.

To start with, let us consider a variation where we define the cost function $F$ using the operator norm $\|\cdot\|_\infty$ rather than the Frobenius norm squared. $\|X\|_\infty$ can be obtained by minimizing $\lambda$ subject to $-\lambda\le X\le \lambda$. Thus, minimizing $F$ amounts to solving the SDP
\begin{align*}
\operatorname*{minimize}_{\{c_a\},P,\lambda}\ &\lambda\\
\textrm{subject to } & 
    -\lambda \le  \sum_a c_a O_a - 
            \sum_{\alpha, \beta } P_{\alpha\beta} \ketbra{\varphi_\alpha}{\varphi_\beta}  \le \lambda 
\\
& P\ge 0\ .
\end{align*}
In case the cost function is defined using the Frobenius norm, as in Eq.~\eqref{eigvprob2}, one can rewrite $\|X\|_2$ as the minimum of $\lambda$ subject to
\[
\begin{pmatrix} \lambda\openone & X \\ X^\dagger & \lambda\openone
\end{pmatrix} \ge 0\ ,
\]
to yet again re-express the minimization of $F$ as an SDP.

\section{Benchmarks}

Let us now benchmark our method with two well-studied models: 
The 1D AKLT model and the 
Toric Code on the square lattice.

\subsection{The AKLT state on a chain}

\begin{figure}[b]
\center
 \includegraphics[scale=0.4]{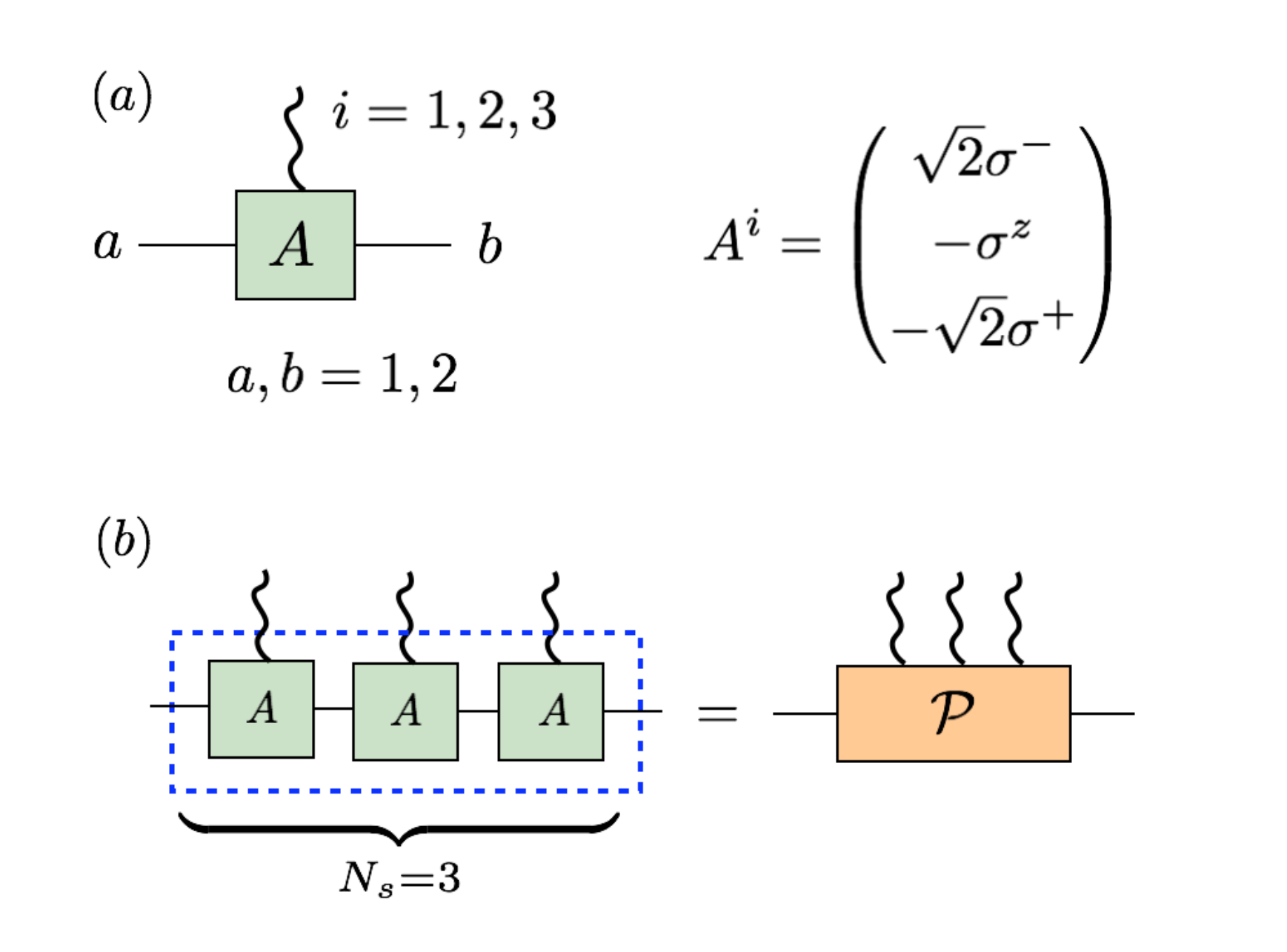} 
\caption{(a) Tensor Network representation of the AKLT state. (b)~Blocking $3$ sites of the MPS tensors yields the map $\mathcal{P}$, starting point of the parent Hamiltonian construction. In this case the number of blocked sites $N_s$ is larger than the minimum $N_s$ required to build the most local parent Hamiltonian Eq.~\eqref{ham_aklt}. }
\label{fig:aklt_1} 
\end{figure}

The AKLT state~\cite{affleck:aklt-prl,affleck:aklt-cmp} is constructed by placing spin-$1/2$ singlets on a chain and projecting every two adjacent spin-$1/2$'s to the joint spin-$1$ (symmetric) subspace. The resulting spin-$1$ chain is rotationally (i.e.\ $\mathrm{SO(3)}$) invariant and can be described as an MPS with the tensors given in Fig.~\ref{fig:aklt_1}a. 
This MPS is normal and becomes injective upon blocking $L=2$ sites. Thus, the elementary proof technique of inverting $\mathcal P$ on the injective block results in a parent Hamiltonian acting on $N_s=4$ sites, while the more refined techniques allow to prove well-behavedness of the parent Hamiltonian on $N_s=3$ sites. 
On the other hand, it is well known that the AKLT state is the unique ground state of the two-body Hamiltonian
\be
H = \sum_{i=1}^L\left[\tfrac12\vec{S}_i \cdot \vec{S}_{i+1} + \tfrac16 ( \vec{S}_i \cdot \vec{S}_{i+1} )^2+ \tfrac13 \right]\ ,
\label{ham_aklt}
\ee
where $\vec{S} = (S^x,S^y,S^z)$ is the spin-$1$ representation of $\mathrm{SU}(2)$. Indeed, the Hamiltonian \eqref{ham_aklt} can be obtained by applying the parent Hamiltonian construction to two sites. Once one has obtained a guess for a two-body Hamiltonian in this way, it is straightforward to verify that it is indeed a well-behaved parent Hamiltonian, by checking that its ground space on three (or four) sites is just the same as that of the the $3$-site (or $4$-site) parent Hamiltonian and thus, it has a unique ground state and a gap.

\begin{figure}
\center
\includegraphics[scale=0.37]{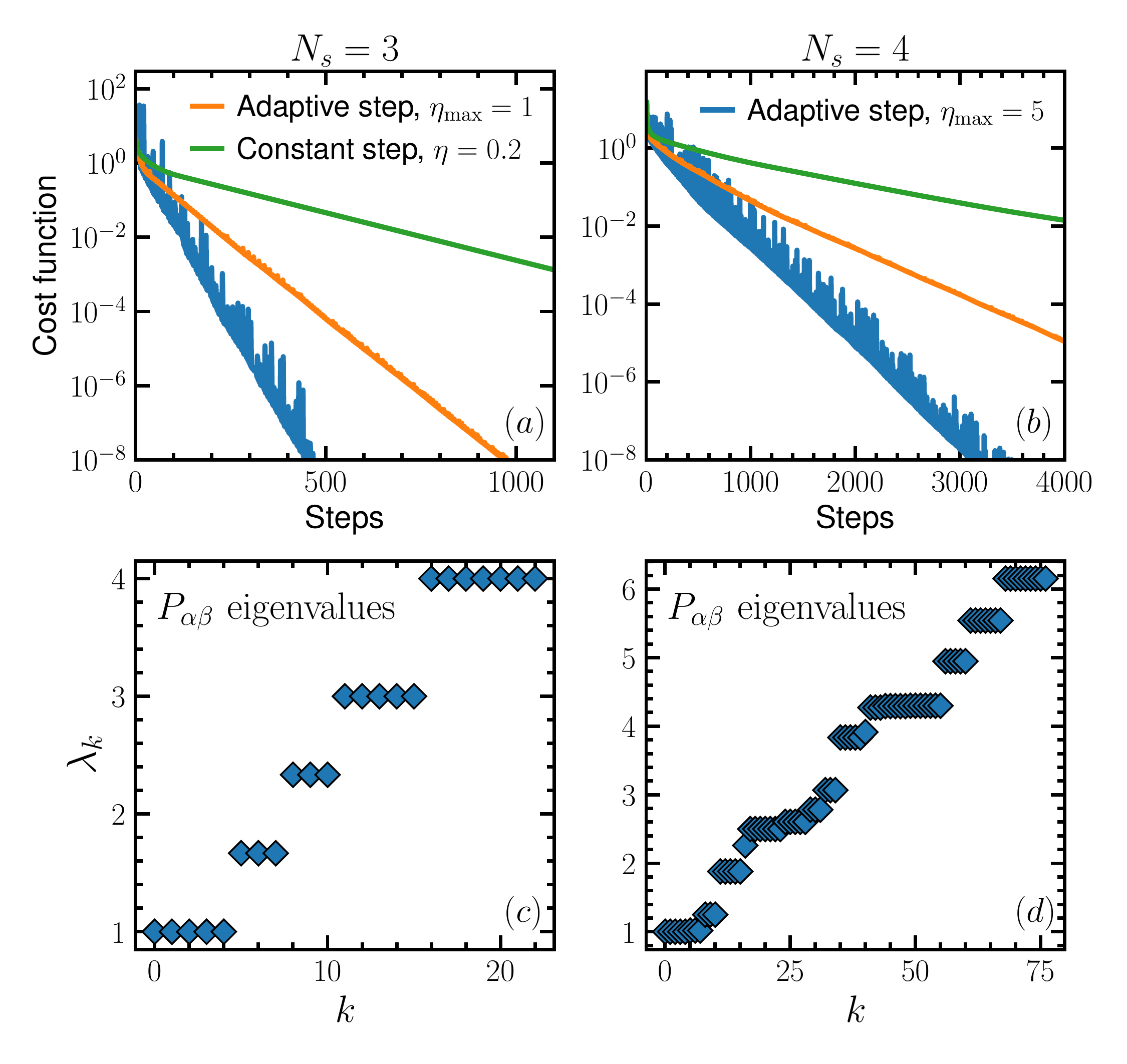}
\caption{(a),(b) Cost function $F$, Eq.~\eqref{eigvprob2}, during the minimization. The adaptive step-gradient descent procedure converges much faster ($\sim 200$ and $1500$ steps to reach $F \sim 10^{-12}$ for $N_s = 3$ and $N_s = 4$, respectively) than the constant step one ($\eta= 0.5$). (c),(d) Eigenvalues of the matrix $P_{\alpha \beta}$ at convergence ($F<10^{-12}$) for $N_s=3,4$. The result shows that the locality-optimized Hamiltonian, acting on $N_s$ sites, is not the 
projector onto $( \mathrm{Im}\,\mathcal{P})^\perp$.     }
\label{fig:aklt_2} 
\end{figure}

For this approach, however, a prior guess for a suitable $2$-body Hamiltonian is required. We will now demonstrate that our method can be used to break up the canonical $3$-body or $4$-body AKLT parent Hamiltonian into 
the $2$-body Hamiltonian \eqref{ham_aklt} \emph{without} any such prior knowledge.
Since we aim to decompose the Hamiltonian into $\mathrm{SU}(2)$-invariant $2$-body nearest neighbor terms, 
we only need to include nearest-neighbor Heisenberg interactions in the operator basis $\{O_a\}$; for instance, with $N_s=3$ the operator basis is $\{ \mathrm{Id}, \vec{S}_1 \cdot \vec{S}_2 ,  ( \vec{S}_1 \cdot \vec{S}_2)^2  , \vec{S}_2 \cdot \vec{S}_3  , (\vec{S}_2 \cdot \vec{S}_3)^2 \}$. Fig.~\ref{fig:aklt_2} shows the results of the minimization procedure. In Fig.~\ref{fig:aklt_2}a and Fig.~\ref{fig:aklt_2}b we plot the cost function during the (projected) gradient descent optimization for $N_s = 3$ and $N_s = 4$, respectively. We compare a constant step $\eta = 0.2$ (blue line) and an adaptive step $\eta = \mathrm{Min} \left\{ \langle \delta X_n , \delta X_n  \rangle /  \langle \delta X_n , \delta G_n \rangle , \eta_\mathrm{max} \right\}$ (orange and blue lines)~\footnote{A maximum step size $\eta_\mathrm{max}$ is necessary to ensure the stability of the optimization. The step size $\eta=0.2$ is the maximum size yielding a stable constant step optimization. }.  Fig.~\ref{fig:aklt_2}a and Fig.~\ref{fig:aklt_2}b demonstrate that the matrix $P_{\alpha\beta}$ obtained, and thus the Hamiltonian density, is not a projector on the original $N_s=3,4$ sites. 
Rather, the Hamiltonian obtained is equal to 
Eq.~\eqref{ham_aklt} within numerical precision.

\subsection{The toric code on the square lattice\label{sec:tcode}}

\begin{figure}
 \includegraphics[scale=0.5]{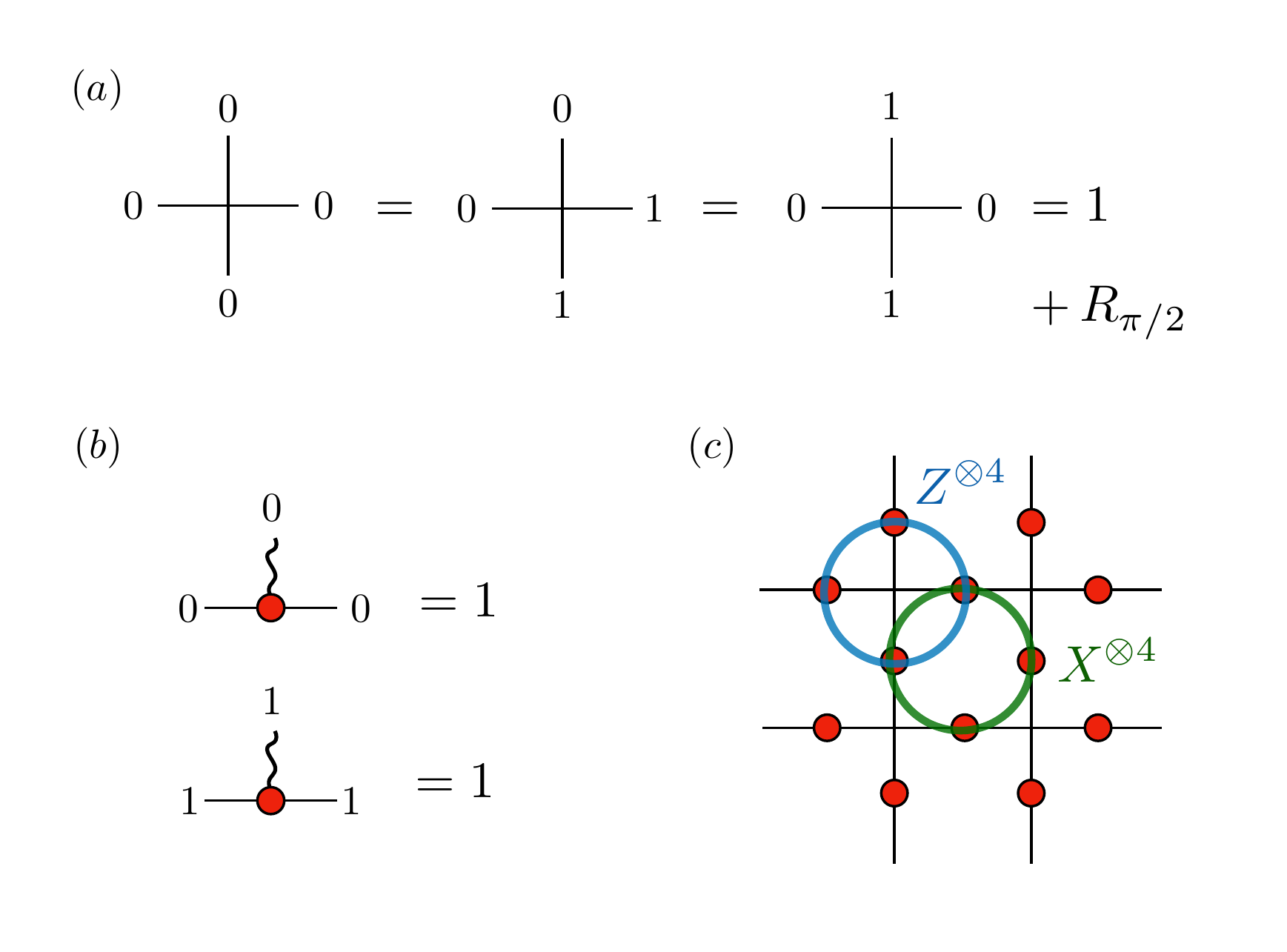} 
\caption{(a) Vertex tensors providing the PEPS representation of the toric code ground state. The virtual index $1$ signals the presence of a loop entering the vertex. $R_{\pi/2}$ stands for all the possible $\pi/2$ rotations of the vertex tensors. (b) Local projectors that map diagonally the virtual space of the vertex tensors to the physical space of a site on the tilted square lattice. (c) $N_s=12$ sites patch used for the parent Hamiltonian construction. Physical legs are omitted in the drawing. The linear map $\mathcal{P}$ obtained from this patch maps the virtual space $(\mathbb{C^2})^{\otimes 8}$ to the physical space $(\mathbb{C^2})^{\otimes 12}$. Plaquette terms are product of $X$s on the green line, cross terms are product of $Z$s on the blue line. }
\label{fig:toric_1} 
\end{figure}

Let us now turn to a paradigmatic two-dimensional model which exhibits topological order: Kitaev's Toric Code~\cite{kitaev2003}.
Consider an arbitrary lattice, and assign a qubit $\{\ket0,\ket1\}$ to every edge of the lattice. The Toric Code wavefunction is then given by the equal weight superposition of all configurations which obey a $\mathbb Z_2$ Gauss law around every vertex, i.e., there are an even number of $\ket1$ states adjacent to every vertex. Equivalently, this amounts to saying that the wavefunction is an equal weight superposition of all closed loop configurations on the lattice, where the states $\ket0$ and $\ket1$ correspond to the no-loop and loop state, respectively. In the following, we focus on the square lattice. A tensor network representation of the Toric Code can be constructed by using two types of tensors: One vertex tensor which only carries virtual indices and which enforces the $\mathbb Z_2$ Gauss law (see Fig.~\ref{fig:toric_1}a) and an edge tensor which is described by a Kronecker delta and which ``copies'' the virtual degree of freedom to a physical qubit (see Fig.~\ref{fig:toric_1}b); these tensors are arranged to a tensor network as shown in Fig.~\ref{fig:toric_1}c. 

Conceptually, a parent Hamiltonian constructed on a region $R$ (with edge tensors at its boundary) ensures that for any given edge configuration, all bulk loop configurations get the same weight. It is easy to see that such a Hamiltonian will enforce the closed loop constraint, as well as enforce an equal amplitude for all loop configurations which can be coupled by local moves; this gives rise to a $4$-fold degeneracy on the torus, with sectors labeled by the parity of loops around the torus in either direction~\cite{kitaev2003,schuch:peps-sym}.
To construct a parent Hamiltonian with controlled ground space degeneracy, we start by noting 
that the map $\mathcal P$ given by 
a vertex tensor surrounded by four edge tensors is injective~\footnote{In principle, three edge tensors are sufficient: This results in a slightly smaller parent Hamiltonian (some edge sites -- but not all! -- at its boundary can be omitted in Fig.~\ref{fig:toric_1}c), which however breaks the lattice symmetry. We thus choose to work with the given $4$-site injective patch.}. A resulting canonical parent Hamiltonian can then be constructed on the $12$-sites patch shown in Fig.~\ref{fig:toric_1}c; it is straightforward to prove that it satisfies the intersection property by using the refined ``invert and grow back'' techniques discussed e.g.\ in Refs.~\cite{schuch:rvb-kagome,cirac:tn-review-2021}, and thus has the correct ground space structure.

\begin{figure}
\center
\includegraphics[scale=0.37]{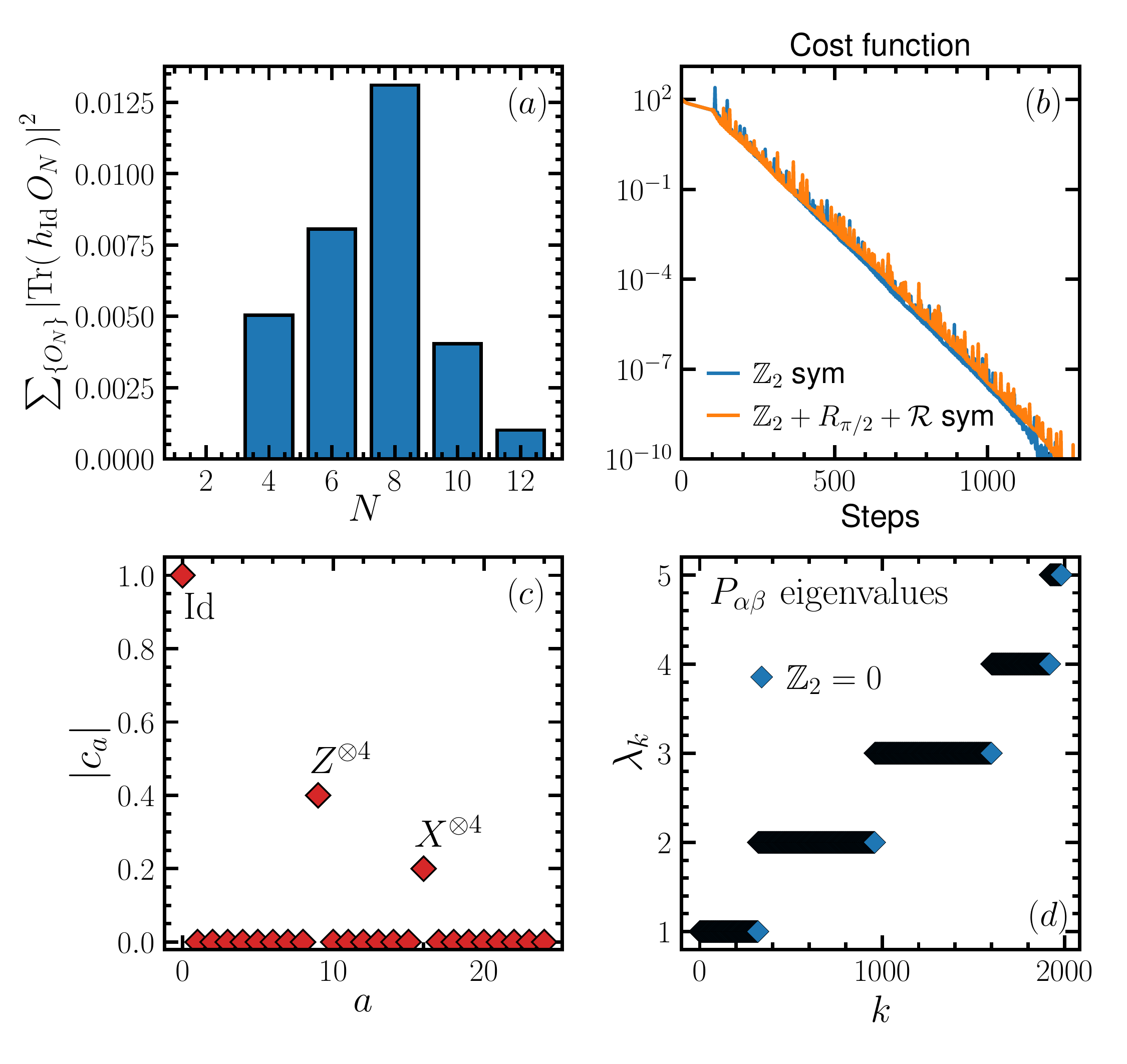}
\caption{(a) 
Decomposition of the projector-valued parent Hamiltonian $h_\mathrm{Id}$ of the Toric Code in terms of $N$-body Pauli products; the plot shows the total weight of all $N$-body terms. Terms with all possible weights are needed in the decomposition.
(b) Cost function during the minimization, using the adaptive step. Symmetric and non-symmetric versions of the algorithm achieve convergence in the same number of steps, but the symmetric version reduces significantly the computation cost. (c) Coefficients of the operators $\{O_a\}$ in the optimal decomposition at convergence. When only cross and plaquette terms (made of $X$,$Y$ or $Z$) are included in the basis, the algorithm produces the expected result within machine precision. (d) Eigenvalues of the matrix $P_{\alpha \beta}$ at convergence, divided in the two physical $\mathbb{Z}_2$ symmetry sectors. The most local Hamiltonian is different from the projector on $( \mathrm{Im}\, \mathcal{P} )^\perp $. }
\label{fig:toric_2} 
\end{figure}
 
In the following, we apply our method to analyze how one can break down the canonical parent Hamiltonian acting on patches of $N_s=12$ sites into simpler terms, where one goal is to check
whether our optimization method does indeed return the Toric Code Hamiltonian
\be 
H = - \sum_v \prod_{i \in v} Z_i - \sum_p \prod_{i \in p} X_i\ ,
\label{eq:toric_ham}
\ee
which in known to have the Toric Code wavefunction as its unique ground state (up to topological degeneracy)~\cite{kitaev2003}.
 Here,
the terms in the first sum act on the four qubits surrounding any given vertex $v$, and those in the second sum on the four qubits surrounding any given plaquette $p$ of the square lattice, as shown in Fig.~\ref{fig:toric_1}c, where $X$ and $Z$ are Pauli matrices. 

To this end, we start from the $N_s=12$ site patch depicted in Fig.~\ref{fig:toric_1}c, for which  
the dimension of $(\mathrm{Im}\,\P)^\perp$ is $3968$.
Fig.~\ref{fig:toric_2}a shows the 
decomposition
of the projector-valued Hamiltonian $h_{\mathrm{Id}}$ (i.e. with $P=\mathrm{Id}$ in Eq.~\eqref{eq:h-sum-P})
in an operator basis made of all the possible products of Pauli matrices on the $12$ sites, grouped by the number $N_{{sites}}$ of non-trivial Pauli operators: We observe that $h_{\mathrm{Id}}$ contains terms all the way up to $N_{{sites}}=12$-body terms, and thus, optimization over $P$ is required to obtain a simpler decomposition of the parent Hamiltonian $h$ into Pauli products.

We now apply the algorithm of Sec.~\ref{ph2_2} starting from an operator basis with all possible products of Pauli matrices on $4$-site crosses and plaquettes (see Fig.~\ref{fig:toric_1}c) that respect the physical symmetries of the TN state on the chosen patch: Specifically, we used the $\mathbb{Z}_2$ symmetry generated by $\prod^{N_s}_{i=1} Z_i$, the four-fold $\pi/2$-rotation 
$R_{\pi/2}$ 
around the center of the patch,
and the reflection 
$\mathcal R$ w.r.t.\ the horizontal axis that goes through the center of the patch. This basis contains $25$ operators (including the identity), and the total number of variational parameters for the non-symmetric version of the algorithm would be $25 + 3968^2 \simeq 1.6 \cdot 10^7$. 
Given the aforementioned symmetries, we can apply the symmetric version of the algorithm to reduce the number of parameters. In particular, we consider the algorithm where we use the global $\mathbb{Z}_2$ symmetry only, which reduces the dimension of the variational space down to $25 + 2 \cdot (3968/2)^2 \simeq 7.9 \cdot 10^6$, and the fully symmetric version where we exploit $\mathbb{Z}_2$, $R_{\pi/2}$, and $\mathcal{R}$ symmetries, yielding a variational space dimension of $1466797 \simeq 1.5 \cdot 10^6$, and compare their performance.
The cost function during the minimization is plotted in Fig.~\ref{fig:toric_2}b. Both versions of the algorithm converge to the minimum in $\sim 1000$ steps. However, it is computationally much cheaper for the most symmetric version to perform a single step, reducing the CPU time by a factor proportional to the ratio between the variational space dimensions without and with symmetries. Although in this example, exploiting the full symmetry group of the patch is not indispensable, it will be crucial in the next example, where even a single optimization step would be prohibitive without it.

Considering the optimum found by the algorithm, we find that our method indeed yields a parent Hamiltonian identical to the one in Eq.~\eqref{eq:toric_ham}, except for a different relative weight of the two types of terms (this is possible as the terms commute, and the Hamiltonian is frustration-free, i.e., the ground state minimizes each of the four-body terms individually). In Fig.~\ref{fig:toric_2}c we show the coefficients of the Hamiltonian density on the $12$-site patch at convergence. In Fig.~\ref{fig:toric_2}d we plot the eigenvalues of the optimized parent Hamiltonian $h$ on the $12$-site patch, demonstrating that the most local $h$ is indeed not a projector.

\section{The SU(2) Resonating Valence Bond state on the kagome lattice}

In the following, we apply our method to find an optimally local parent Hamiltonian for the paradigmatic Resonating Valence Bond state on the kagome lattice, which is a prime example of a topological spin liquid. For this model, the simplest hitherto known parent Hamiltonian was a general operator acting on a whole kagome star, that is, $N_s=12$ sites. Applying our method, we arrive at a much simpler Hamiltonian, where each term is a product of at most four Heisenberg interactions. This represents a significant simplification over the original $12$-body interaction and demonstrates the power of our method.

Resonating Valence Bond (RVB) states are constructed as equal-weight superpositions of all possible singlet coverings between nearest neighbors on a given 2D lattice. In quantum dimer models, singlets are replaced by orthogonal dimers, facilitating their analysis. On frustrated lattices, such dimer models have been known for a long time to be 
simple representatives of topologically ordered phases~\cite{Moessner2011}; in particular, the dimer model on the kagome lattice is a topological $\mathbb Z_2$ fixed point model, locally unitarily equivalent to the toric code. When orthogonal dimer coverings are replaced by spin-1/2 singlets, one obtains the RVB state, which is a good candidate 
for describing the physics of frustrated magnets. On the kagome lattice, it was shown to be in the same $\mathbb{Z}_2$ spin liquid phase as the kagome dimer model~\cite{schuch:rvb-kagome}, and demonstrated to be even more stable against perturbation~\cite{mohsin2020}. Starting from the SU(2) singlets RVB state, simple ansatze for the ground state wave functions of physically relevant models have been devised~\cite{mohsin2020_2}. 
RVB states have a natural PEPS representation~\cite{verstraete:comp-power-of-peps,schuch:peps-sym} with a $\mathbb Z_2$ entanglement symmetry; they become $\mathbb Z_2$-injective upon blocking and thus, their canonical parent Hamiltonians exhibit a four-fold degenerate ground space on the torus, as required for a $\mathbb Z_2$ topological spin liquid.
However, the hitherto known parent Hamiltonians for the kagome RVB state are rather complicated: The simplest parent Hamiltonian which has been obtained using canonical techniques is constructed on two overlapping stars, that is, $19$ sites~\cite{schuch:rvb-kagome}; later, it has been shown by brute-force numerical checking that the parent Hamiltonian constructed from the map $\mathcal P$ on a single star, that is, $N_s=12$ sites, has the same ground space as the $19$-site two-star Hamiltonian when applied on both of the stars~\cite{zhou:rvb-parent-onestar}. 
In the following, we will apply our algorithm to analyze whether, and how, this one-star Hamiltonian can be broken down into simpler terms.

\begin{figure}
\center
\includegraphics[scale=0.5]{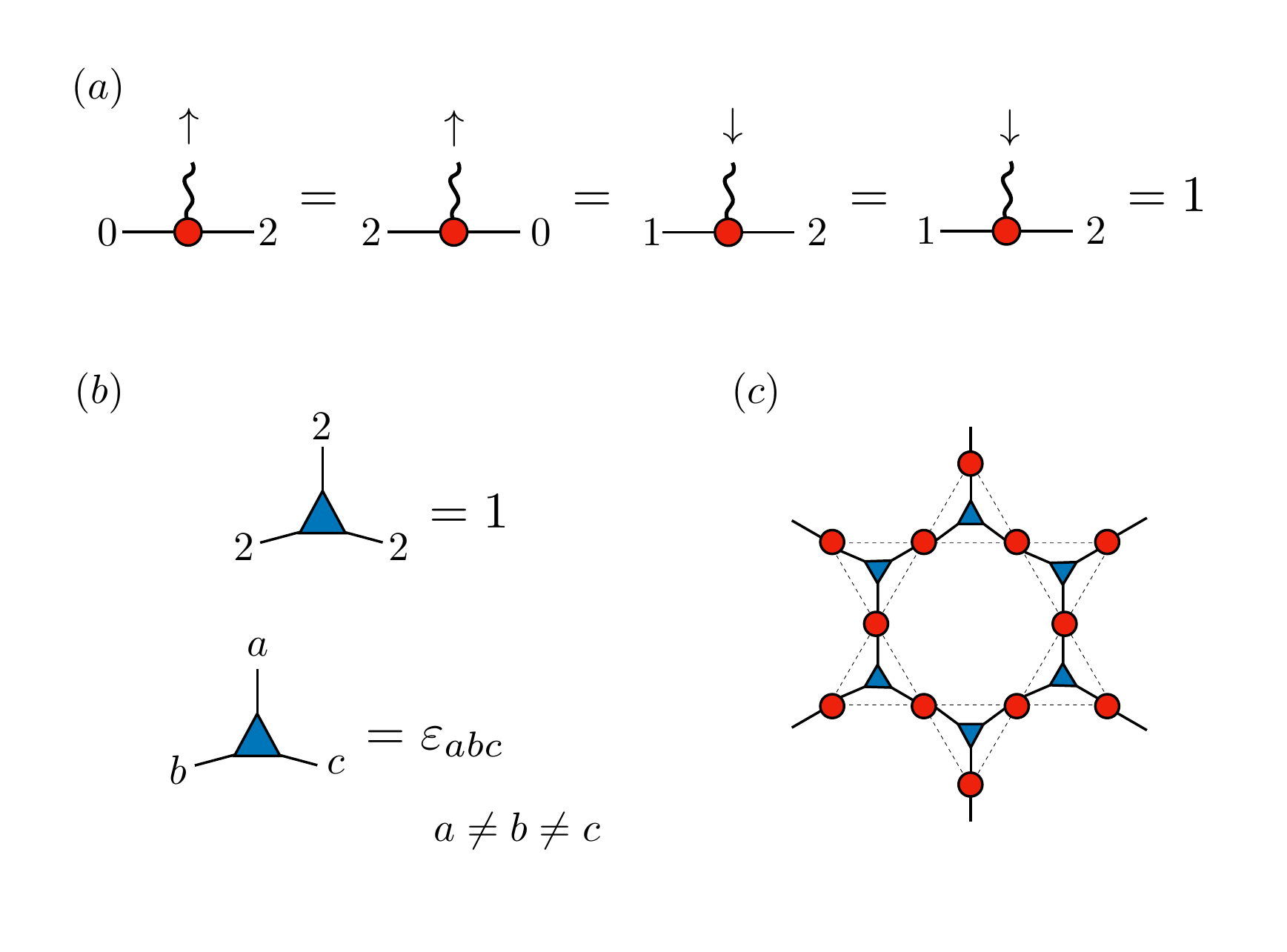} 
\caption{TN representation of the spin-$1/2$ RVB state.
The on-site tensors (a) and triangle plaquette tensors (b) -- only non-zero entries are shown -- are placed and contracted on the kagome lattice as shown in (c). The states $0$ and $1$ carry the spin-$1/2$ degree of freedom, while the state $2$ signals the absence of a spin-$1/2$. The on-site tensor selects either the left or right singlet and maps it to the physical spin-$1/2$; the triangle plaquette tensor can either hold no singlet ($222$ configuration), or exactly one singlet ($\varepsilon_{abc}$).
Up- and downward-pointing blue tensors are related by rotation.}
\label{fig:rvb-tnrep} 
\end{figure}

We use the TN representation of the SU(2) RVB state introduced in Ref.~\cite{schuch:rvb-kagome}, which is given in Fig.~\ref{fig:rvb-tnrep}. Fig.~\ref{fig:rvb-tnrep}c shows the patch of $12$ sites that we consider in what follows, where the virtual space consists of $6$ bonds with dimension $D=3$. The PEPS is $\mathbb{Z}_2$-injective, with the virtual symmetry generator $g = \mathrm{diag} ( 1 , 1 , -1 )$. 
Since the dimension of $(\mathrm{Im}\,\P)^\perp$ is $3731$, the number of variational parameters which arise from the matrix $P$ (Eq.~\eqref{eq:h-sum-P}) would be more than $10^7$. As this is prohibitive, we will need to use the symmetries of the system at hand.

Let us start with the symmetries of the parent Hamiltonian $h$, that is, the basis $\{\ket{\varphi_\alpha}\}$ on which $h$ is supported and the associated positive matrix $P>0$, $h=\sum P_{\alpha\beta} \ket{\varphi_\alpha}\bra{\varphi_\beta}$. The map $\mathcal P$ of the one-star tensor network in Fig.~\ref{fig:rvb-tnrep}c (which maps the virtual to the physical system) commutes with both $\mathrm{SU}(2)$ and a $60\degree$ rotation; by choosing boundary conditions with well-defined quantum numbers on the virtual system, we can thus ensure that we obtain $\ket{\varphi_\alpha}$ with well-defined $\mathrm{SU}(2)$ and angular momentum quantum numbers. In addition, we can ensure that the $\ket{\varphi_\alpha}$ transform nicely under reflection about the vertical axis: Such a reflection changes $\varepsilon_{abc}\to-\varepsilon_{abc}$ in Fig.~\ref{fig:rvb-tnrep}b, that is, the state acquires a minus sign for each blue tensor in the $\varepsilon_{abc}$ configuration. The number of those tensors equals the number of triangles which hold a singlet (or dimer), and the number of singlets inside the star is determined by a simple counting argument from the number of spin-$1/2$ states at the virtual boundary (i.e., boundary configurations $\{\ket{0},
\ket{1}\}$), or more precisely this number modulo $4$. This quantum number of the boundary condition commutes with $\mathrm{SU}(2)$ and angular momentum, and thus, we can obtain a basis of $\ket{\varphi_\alpha}$ labeled by $\mathrm{SU}(2)$, rotation, and reflection quantum numbers.
By exploiting these symmetries, as described in Sec.~\ref{sec:method-symmetries},
we are able to break down the $P_{\alpha \beta}$ variational matrix into blocks, yielding $20931$ parameters.

\begin{figure}
\center
\includegraphics[scale=0.7]{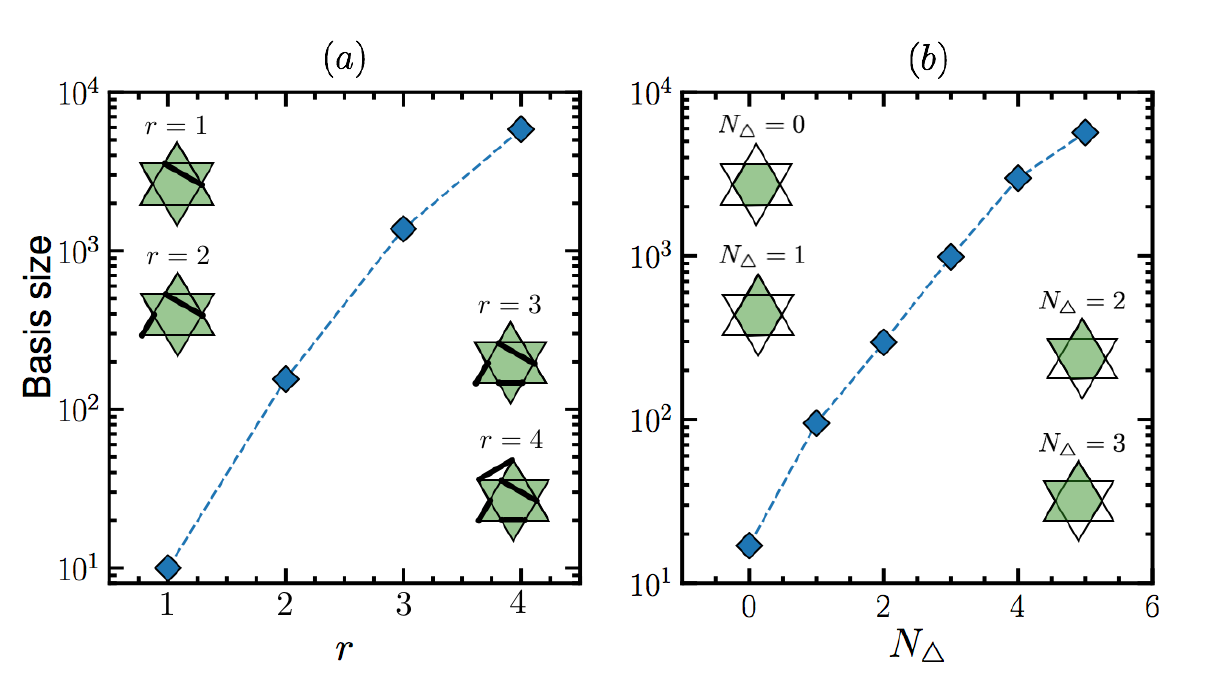}
\caption{Operator basis dimension for the two cases considered here. (a) The basis contains all products of up to $r$ Heisenberg interactions. (b) 
The basis contains all products of up to $r=4$ Heisenberg interactions, restricted to the spins on the central hexagon and up to $N_\triangle$ continguous outer spins.
In both cases, the operators are symmetrized w.r.t.\ the six-fold rotation and two-fold reflection symmetry of the star, see text.}
\label{fig:rvb-basissets}   
\end{figure}

We exploit the same symmetries to select the operators to be included in the basis $\{O_a\}$. As we show in the Appendix, every  $\mathrm{SU}(2)$-invariant Hamiltonians $H$ on $(\mathbb C^2)^{\otimes N}$ can be decomposed as $H=\sum w_a K_a$ in a basis of $\mathrm{SU}(2)$-invariant Hamiltonians $\{K_a\}$, where each $K_a$ is a 
non-overlapping product of only two types of terms: 
Heisenberg interactions  $\vec S^i\cdot\vec S^j$, and ``chiral'' terms of the form  $\vec S^i\cdot(\vec S^j\times \vec S^k)$; in addition, at most one such chiral term is needed. 
We use this basis of $\mathrm{SU}(2)$-invariant Hamiltonians to build the basis of operators $\{O_a\}$ for our parent Hamiltonian. In this process, we make use of further symmetries: First, we observe that the RVB wavefunction is real, and thus, the parent Hamiltonian can be chosen real as well, $h=\bar h$ (more specifically, we can replace any parent Hamiltonian $h$ by $(h+\bar h)/2$, which changes neither the ground space nor the spectrum). Since 
$\vec S^1\cdot(\vec S^2\times \vec S^3)$ is purely imaginary (as it is a sum of terms with one $S_y$ each) and it appears at most once, this implies that we can omit it altogether, and the basis $\{O_a\}$ can be chosen to be spanned solely by non-overlapping products of Heisenberg terms. Further, we impose that the basis $\{O_a\}$ has the same lattice symmetries as used for the $\{\ket{\varphi_{\alpha}}\}$, i.e.\ reflection about the vertical axis and rotation by $60\degree$.

\begin{figure}
\hspace*{-3mm}
 \includegraphics[scale=0.35]{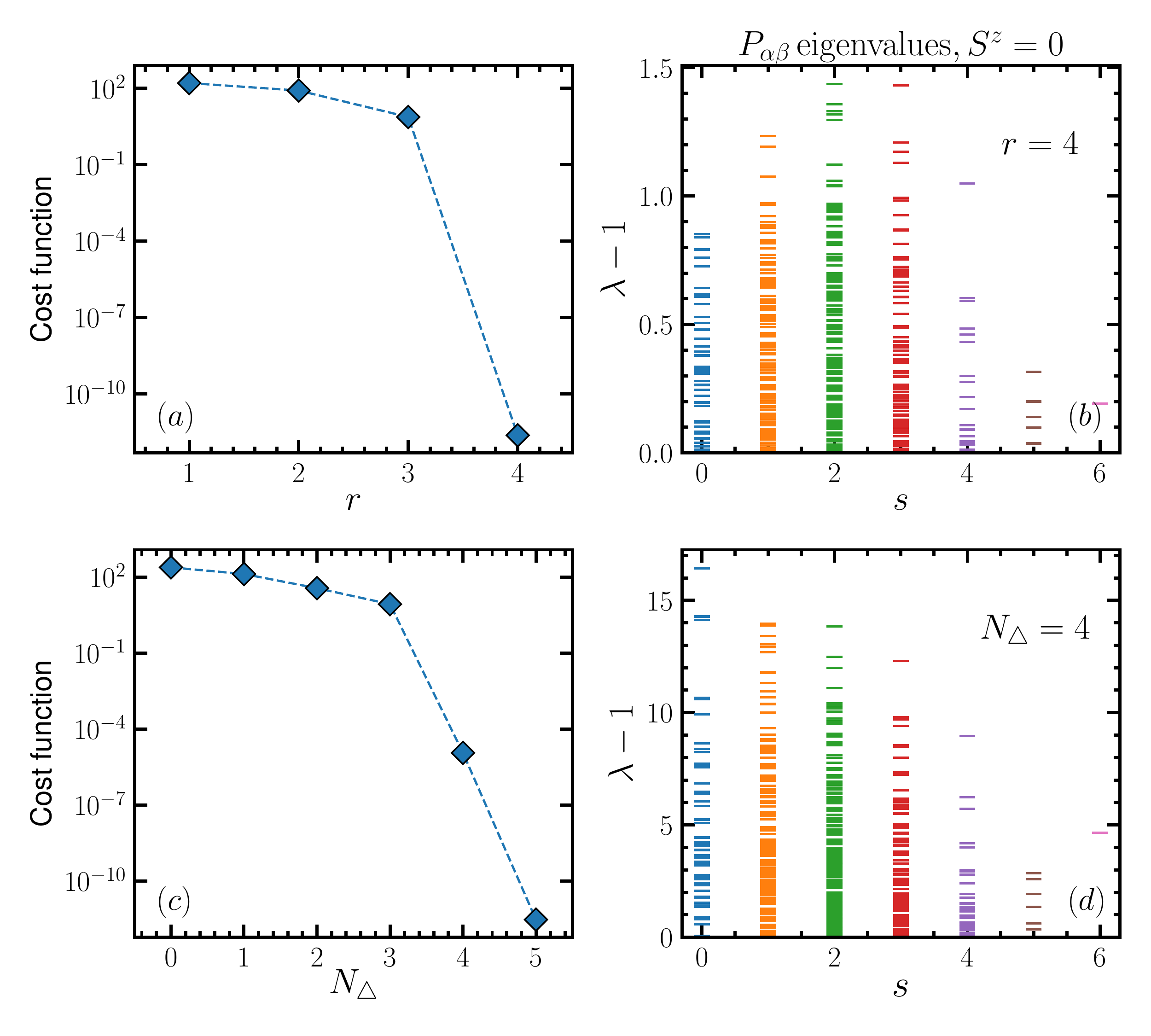} 
\caption{(a,c) Cost function $F$, Eq.~\eqref{eigvprob2}, at the optimum 
as a function of $r$ and $N_\triangle$, respectively (see Fig.~\ref{fig:rvb-basissets}).
A machine-precision approximation is obtained for $r=4$ and $N_\triangle=5$, respectively, but $N_\triangle=4$ already provides a fairly accurate approximation. 
(b,d) Eigenvalues of $P_{\alpha \beta}$ in the $S^z = 0$ sector, vs.\ the value $s(s+1)$ of the total spin $S^2$, for $r=4$ and $N_\triangle$, respectively, showing that the optimal solution is not a projector. 
}
\label{fig:rvb-optimization}  
\end{figure}

We thus have that each operator $O_a$ is a symmetrized product of $1\le r\le 6$ Heisenberg interactions. As we are interested in finding the most local parent Hamiltonian, we apply our method to a growing sequence of basis sets $\mathcal O_r = \{O_a\}$, where a given $\mathcal O_r$ consists of all operators with \emph{up to} $r$ Heisenberg interaction (see Fig.~\ref{fig:rvb-basissets}a for the dimension of those basis sets, going up to about $6000$). Fig.~\ref{fig:rvb-optimization}a shows the results obtained by our algorithm when approximating the parent Hamiltonian with the basis set $\mathcal O_r$, for $r=1,\dots,4$. We find that while for $r\le 3$, the parent Hamiltonian cannot be faithfully approximated, the basis set $\mathcal O_4$, which contains products of up to $r=4$ Heisenberg terms, provides an approximation of the exact parent Hamiltonian up to machine precision~\footnote{Depending on the basis size the algorithm may take several thousands of steps to converge, and it does not always find an exact parent Hamiltonian. This is signaled by the fact that at convergence -- we monitor the status of convergence by computing the norm displacement vector at consecutive steps $\epsilon = || X_n - X_{n-1} || $, and stop the algorithm when $\epsilon < 10^{-10}$ -- the cost function is not vanishing within numerical precision.}.

In order to investigate whether it is possible to further simplify the Hamiltonian, we analyze the effect of restricting $\mathcal O_r$ to more local Heisenberg terms; specifically, we define $\mathcal O_4(N_\triangle)$ to contain all products of up to $4$ Heisenberg terms which act on the six central spins and up to $N_\triangle$ adjacent spins at the tips of the star, as shown in Fig.~\ref{fig:rvb-basissets}b (which also provides the basis size). Fig.~\ref{fig:rvb-optimization}c shows the results on the accuracy of the approximation of the parent Hamiltonian using the basis set $\mathcal O_4(N_\triangle)$: While for $N_\triangle\le 3$, the parent Hamiltonian is not well approximated, we can reproduce the Hamiltonian up to machine precision for $N_\triangle=5$. The case $N_\triangle=4$ lies in between those regimes, as it provides a very good but not perfect approximation of the parent Hamiltonian, with an error of about $10^{-5}$ in Frobenius norm squared.

\begin{figure}
\center
\includegraphics[scale=0.7]{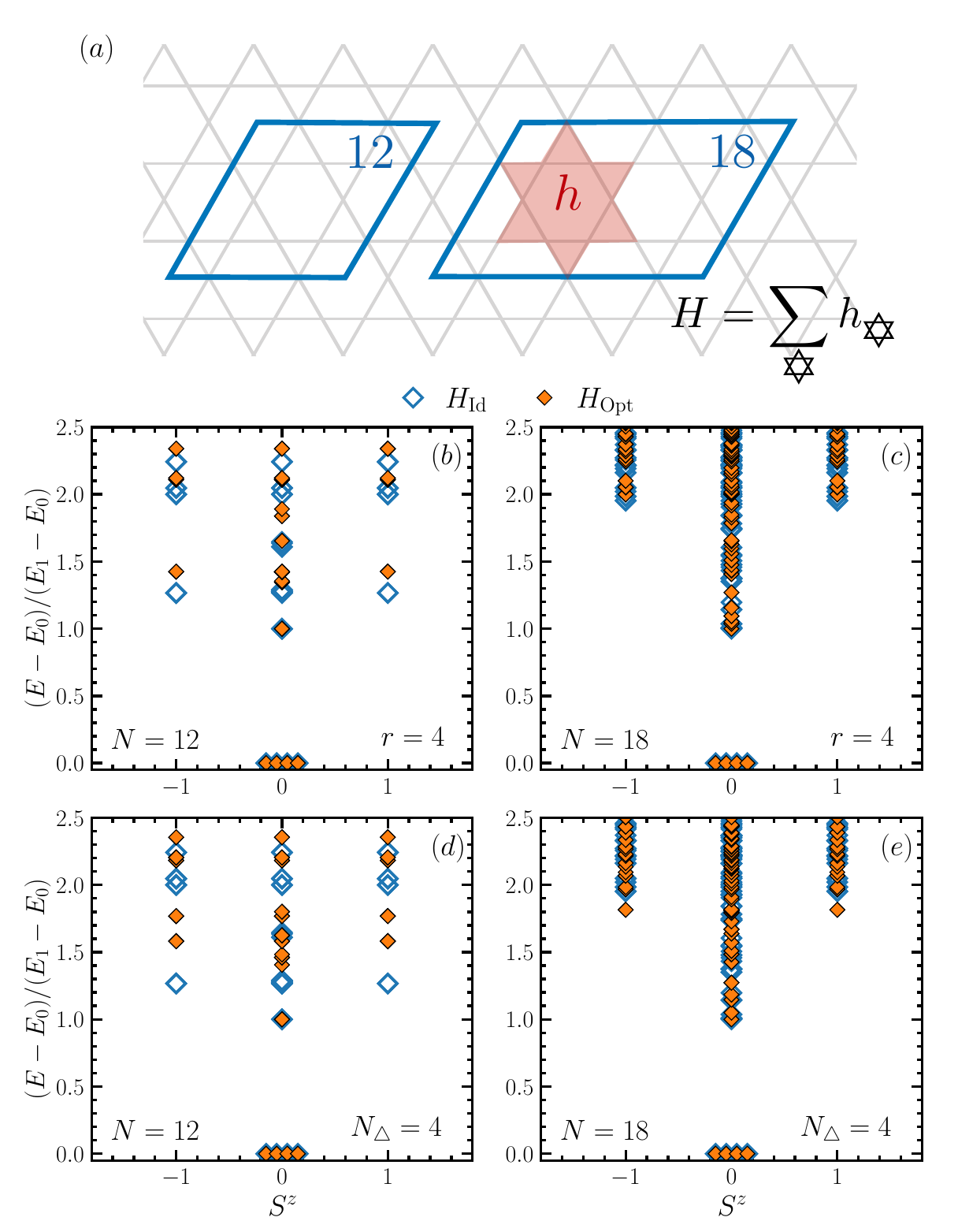}
\caption{(a) Periodic clusters employed for the computation of the spectrum of the parent Hamiltonian. The latter is obtained as a sum of the parent Hamiltonian $h$ on all stars of the cluster. (b-e) Comparison between the low energy spectra of the RVB parent Hamiltonian obtained as a projector on $(\mathrm{Im}\,\mathcal{P})^\perp$ (blue empty diamonds) and the optimized Hamiltonian (orange filled diamonds) obtained on a basis of operators with up to four Heisenberg interactions (b,c), and up to four triangles around the hexagon (d,e). Periodic clusters with $N=12$ (b,d) and $N=18$ (c,e) are considered. }
\label{fig:rvb2_2} 
\end{figure}

In order to double-check the correctness of the Hamiltonian found by our algorithm, we performed exact diagonalization on small tori, employing periodic clusters including up to 18 sites on the kagome lattice (see Fig.~\ref{fig:rvb2_2}a). Fig.~\ref{fig:rvb2_2} shows comparisons between the spectrum of the $12$-site projector-valued parent Hamiltonian and the optimized Hamiltonians returned by our method (for $r=4$ and $N_\triangle=4$, respectively), for clusters of $12$ and $18$ sites. The expected topological four-fold degeneracy of the ground space is accurate within $10^{-9}$; this is in agreement with the accurate approximation of the parent Hamiltonian whose ground space has an exact four-fold degeneracy.  On the other hand, as expected, the two spectra already differ for low-lying excited states. 
Similarly, the spectra of the optimized parent Hamiltonians on the full star for $r=4$ and $N_\triangle=4$ 
(Fig.~\ref{fig:rvb-optimization}b and Fig.~\ref{fig:rvb-optimization}d,
 respectively)  are far from the projector-valued parent Hamiltonian.
We also verified that the overlap of the RVB state with the ground-space of the optimized Hamiltonian equals $1$ within numerical accuracy, for clusters up to $N=24$ sites. 

An interesting question is whether one can 
gain physical intuition on the operators that are essential to obtain a good parent Hamiltonian for the RVB state. Unfortunately, we found this to be hard in practice, due to the large number of operators in the basis ($\sim 3000$ at least). Although certain operators contribute far more than others to the final result (e.g., the number of operators $O_a$ with weights $|c_a/c_0| > 0.5$, where $c_0$ is the largest coefficient of the identity operator, are only $10$), all the operators appear to be necessary to give high overlap with the RVB state, and to reproduce the correct ground state degeneracy.

\section{Conclusions}
\label{ph2_4}

In this work, we have presented an algorithm to systematically simplify parent Hamiltonians for tensor network states such as MPS and PEPS.
Specifically, our method allows to decompose a parent Hamiltonian into a sum of terms chosen from any given set of elementary operators, while preserving the ground space as well as the gappedness of the original parent Hamiltonian. A central ingredient is the remaining degree of freedom in the parent Hamiltonian construction: while its local terms have a fixed ground space, they can have an arbitrary excitation spectrum. Our method exploits this degree of freedom to optimize for the parent Hamiltonian which is best approximated by the given basis set of elementary interactions. 
This results in a convex optimization problem which can formally be written as a semidefinite program, and which can thus be solved efficiently by a gradient-based algorithm to find the optimal decomposition. Additionally, our algorithm can be combined with symmetries of the parent Hamiltonian (that is, the symmetries of the tensor network state) by imposing the same symmetries on the basis set, which leads to a significant reduction in computational resources.

We have applied our method to three paradigmatic tensor network models: First, the 1D AKLT model, where we found that our algorithm allows to decompose the canonically constructed $3$- or $4$-body parent Hamiltonians into the well-known $2$-body AKLT Hamiltonian. Next, the Toric Code model, where refined canonical constructions yield a parent Hamiltonian acting on $12$ sites, which our algorithm would break down into the well-known Toric Code Hamiltonian consisting of $4$-body vertex and plaquette terms. 

Finally, we have studied the Resonating Valence Bond (RVB) state on the kagome lattice, which is a prime example of a topological spin liquid, and which possesses a succinct PEPS representation. However, while the RVB state (as a $\mathbb Z_2$-injective PEPS) is the exact $4$-fold degenerate ground state of a local parent Hamiltonian, the smallest canonically constructed parent Hamiltonian still acts on 2 overlapping stars, that is, $19$ spins, which had been shown by brute force to be decomposable as the sum of two one-star, i.e., $12$-body, terms. The application of our algorithms to this $12$-site Hamiltonian results in a significant simplification: We find that the RVB parent Hamiltonian can be decomposed into a sum of interaction terms each of which is a product of no more than four Heisenberg interactions, a notable improvement over the general $12$-body interaction; moreover, the range of those interactions can be further restricted. 

Altogether, this demonstrates the ability of our algorithm to significantly simplify the locality of parent Hamiltonians beyond the abilities of existing proof techniques, and opens up the possibility to identify tensor network models which are ground states of particularly simple parent Hamiltonians.

\begin{acknowledgements}
We acknowledge helpful discussions with 
M.~Dalmonte, H.~Dreyer,  G.~Giudice,  M.~Iqbal,  N.~Pancotti,  D.T.~Stephen,  and F.M.~Surace. 
This work has been supported by the  European Research Council (ERC) under the European Union’s Horizon 2020 research and innovation programme through Grant No.~863476 (ERC-CoG SEQUAM) and 
Grant No.~771891 (ERC-CoG QSIMCORR),
as well as the  Deutsche For\-schungs\-ge\-mein\-schaft (DFG, German Research Foundation) under Germany's Excellence Strategy  (EXC-2111 -- 390814868).
\end{acknowledgements}

\appendix*

\section{Decomposition of SU(2)-invariant Hamiltonians into Heisenberg and chiral 3-body terms}

In this appendix, we show that any $\mathrm{SU}(2)$-invariant Hamiltonian  can be decomposed as a sum of terms, each of which only consists of non-overlapping products of Heisenberg interactions 
$\vec S^1\cdot\vec S^2$
and (at most one) chiral 3-body term
$\vec S^1\cdot(\vec S^2\times \vec S^3)$.

Let us first consider a general (not necessarily hermitian) $\mathrm{SU}(2)$-invariant operator $X$ on  $(\mathbb C^2)^{\otimes N}$.
It is well-known that the space of all operators $X$ such that $[X,u^{\otimes N}]=0$ for all $u\in\mathrm{SU}(2)$ is spanned by the canonical representation $V_\pi$ of the permutation group $\mathfrak S_n\ni \pi$, where $V_\pi$ acts by permuting the tensor components of $(\mathbb C^2)^{\otimes N}$~\cite{simon1996representations}. Any permutation $\pi$ can be expressed as a product of 2-cycles $(i,j)$, i.e., swapping two elements $i$ and $j$. Since $V_{(i,j)} = 
(\vec S^i\cdot\vec S^j + \openone)/2$ (where we normalize the spin operator $\vec S^i = (S^i_x,S^i_y,S^i_z)$ at position $i$ to have eigenvalues $\pm1$), we find that the space of all $X$ is spanned by products of Heisenberg interactions $\vec S^i\cdot\vec S^j = \sum \delta_{ab} S^i_aS^j_b$. Note that this includes cases where the different Heisenberg terms overlap.

We will now show that one can get rid of overlapping Heisenberg terms at the cost of introducing just one additional type of interaction, namely $\vec S^i\cdot(\vec S^j\times\vec S^k) = \sum \varepsilon_{abc} S^i_a S^j_b S^k_c$. To this end, consider the overlapping term 
\begin{align*}
&(\vec S^1\cdot \vec S^2)(\vec S^2\cdot \vec S^3) 
= 
    \sum_{abcd} \delta_{ab} S^1_a S^2_b
        \delta_{cd} S^2_c S^3_d\\
&\quad\stackrel{(*)}{=}
    \sum_{abcde}\delta_{ab}\delta_{cd}i\varepsilon_{bce}S^1_aS^2_eS^3_d
    +\sum_{abcd}\delta_{ab}\delta_{cd}\delta_{bc}S^1_aS^3_d
\\
& \quad = 
    \sum_{bce}i\varepsilon_{bce}S^1_aS^2_eS^3_d
    +\sum_{ad}\delta_{bc}S^1_aS^3_d
\\ 
& \quad = 
    i\:\vec S^1\cdot(\vec S^2\times \vec S^3) + \vec S^1\cdot\vec S^3\ ,
\end{align*}
where in $(*)$, we have used that 
\begin{equation}
\label{eq:spinprod}
    S_aS_b = \sum_c i\varepsilon_{abc} S_c + \delta_{ab} \openone\ .
\end{equation}
We have thus succeeded in rewriting a product of two overlapping Heisenberg interactions as a linear combination of  elementary two- and three-body interactions  
$\vec S^i\cdot\vec S^j$ and $\vec S^i\cdot(\vec S^j\times \vec S^k)$. 
We can now continue this analysis for all possible overlaps of those two types of terms, making use of \eqref{eq:spinprod} and the summation rules for $\delta$ and $\varepsilon$ tensors. Even without carrying out this analysis explicitly, it is easy to see that it allows us to transform arbitrary products of those two- and three-body terms into \emph{non-overlapping} products of the same two types of terms: Every application of \eqref{eq:spinprod} gets rid of one overlap of two terms at one position while introducing an $\varepsilon$, and sums over $\delta$ and $\varepsilon$ tensors always yield linear combinations of products of $\delta$ and $\varepsilon$ tensors with no joint indices.
Note that this also gives a constructive procedure to arrive at such a decomposition.

We thus find that any operator $X$ with $[X,u^{\otimes N}]=0$ can be expressed as a complex linear combination of non-overlapping products of terms 
$\vec S^i\cdot\vec S^j$ and $\vec S^i\cdot(\vec S^j\times \vec S^k)$; since those are all hermitian, they also span the set of all hermitian matrices $H$ with $[H,u^{\otimes N}]=0$ over the real numbers.
An additional simplification can be obtained by observing that a product of two $\varepsilon$ can be replaced by a linear combination of $\delta$'s~\footnote{ $ \varepsilon _{ijk}\varepsilon _{lmn} =\delta _{il} \left( \delta _{jm}\delta _{kn}-\delta _{jn}\delta _{km}\right)-\delta _{im}\left(\delta _{jl}\delta _{kn}-\delta _{jn}\delta _{kl}\right)+\delta _{in}\left(\delta _{jl}\delta _{km}-\delta _{jm}\delta _{kl}\right)$ },
and thus, in the basis we require only products of Heisenberg interactions with at most one three-body term $\vec S^1\cdot(\vec S^2\times \vec S^3)$.

\bibliography{biblio,intro}
\end{document}